\documentclass[10pt, conference, letterpaper]{IEEEtran}
\usepackage{xcolor}
\IEEEoverridecommandlockouts
\usepackage{amsmath,amssymb,amsfonts}
\usepackage{gensymb}
\usepackage{algorithmic}
\usepackage{graphicx}
\usepackage{textcomp}
\usepackage{xcolor}
\usepackage{xspace}
\usepackage{enumitem}
\usepackage{url}
\usepackage{caption}
\usepackage{subcaption}
\usepackage{multirow}
\usepackage[normalem]{ulem}
\def\BibTeX{{\rm B\kern-.05em{\sc i\kern-.025em b}\kern-.08em
    T\kern-.1667em\lower.7ex\hbox{E}\kern-.125emX}}
\graphicspath{ {./images/} }

\newcommand{\ie}{\textit{i.e.,}\xspace}
\newcommand{\eg}{\textit{e.g.,}\xspace}

\newcommand{\vs}{\textit{vs.}\xspace}

\newcommand{\dl}{\textsf{DL}\xspace}
\newcommand{\ul}{\textsf{UL}\xspace}
\newcommand{\ping}{\textsf{ping}\xspace}

\newcommand{\vzw}{\textsf{VZW}\xspace}
\newcommand{\cln}{\textsf{CLN}\xspace}

\begin{document}


\title{A Measurement Study of the Impact of Adjacent Channel Interference between C-band and CBRS  \thanks{\hrule \vspace{4pt} This research is supported in part by the National Science Foundation under grant number CNS-229387.}}

\author{\IEEEauthorblockN{
Muhammad Iqbal Rochman\IEEEauthorrefmark{1},
Vanlin Sathya\IEEEauthorrefmark{2}, 
Monisha Ghosh\IEEEauthorrefmark{3},
Bill Payne\IEEEauthorrefmark{1},
and Mehmet Yavuz\IEEEauthorrefmark{2}}
\IEEEauthorblockA{\IEEEauthorrefmark{1}University of Chicago, 
\IEEEauthorrefmark{2}Celona, Inc.,
\IEEEauthorrefmark{3}University of Notre Dame.\\ 
Email: \IEEEauthorrefmark{1}\{muhiqbalcr,billpayne\}@uchicago.edu, \IEEEauthorrefmark{2}\{vanlin,mehmet\}@celona.io, \IEEEauthorrefmark{3}mghosh3@nd.edu}}


\maketitle

\begin{abstract}

The 3.7 - 3.98 GHz frequency band (also known as C-band) was recently allocated in the US for the deployment of 5G cellular services. 
Prior to this, the lower adjacent band, 3.55 - 3.7 GHz, had been allocated to Citizens Broadband Radio Service (CBRS), where the entire 150 MHz can be used for free by Tier 3 General Authorized Access (GAA) users, but access to the spectrum needs to be authorized by the Spectrum Access System (SAS). GAA users are allowed on a channel only when there are no Tier 1 Incumbents (Navy radars) or Tier 2 Priority Access License (PAL) users in the area. However, since there are no guard bands between GAA and C-band, and both systems employ Time Division Duplexing (TDD) where the uplink/downlink configurations are not synchronized, adjacent channel interference can potentially reduce the performance of both systems. In this paper, we quantify the effect of this mutual interference by performing experiments with a real-world deployment. We observe significant downlink throughput reductions on both systems when two devices are in close proximity to each other, and one is transmitting uplink while the other is transmitting downlink: 60\% for 4G CBRS and 43\% for 5G C-band. We believe that this is the first paper to demonstrate this in a real deployment.
This throughput degradation was reduced when the CBSD changed its channel and operated 20 MHz away from C-band, essentially creating a guard band between the channels. We also demonstrate the improvement in latency under adjacent channel interference by implementing MicroSlicing at the CBSD. Our results indicate that addressing adjacent channel interference due to the lack of guard bands and TDD configuration mismatch is crucial to improving the performance of both CBRS and C-band systems.


\end{abstract}


\section{Introduction}\label{sec:intro}

The increased demands on cellular traffic in terms of range, throughput and latency require the use of frequency bands that combine the favorable propagation characteristics of low-band frequencies ($<$ 1 GHz) 
with the wider bandwidths available in the high-band ($>$ 24 GHz). This has led to increasing swathes of mid-band frequencies being allocated for 5G services. In the US, the three most recent allocations in the mid-band are the 3.7 - 3.98 GHz (C-band)~\cite{lagunas20205g}, the immediately adjacent 3.55 - 3.7 GHz (Citizens Broadband Radio Services, or CBRS)~\cite{jai2021optimal} and the latest allocation of 3.45 - 3.55 GHz for cellular services~\cite{FCC_AMBIT}.

\begin{figure}[t]
\centering
\includegraphics[width=\linewidth]{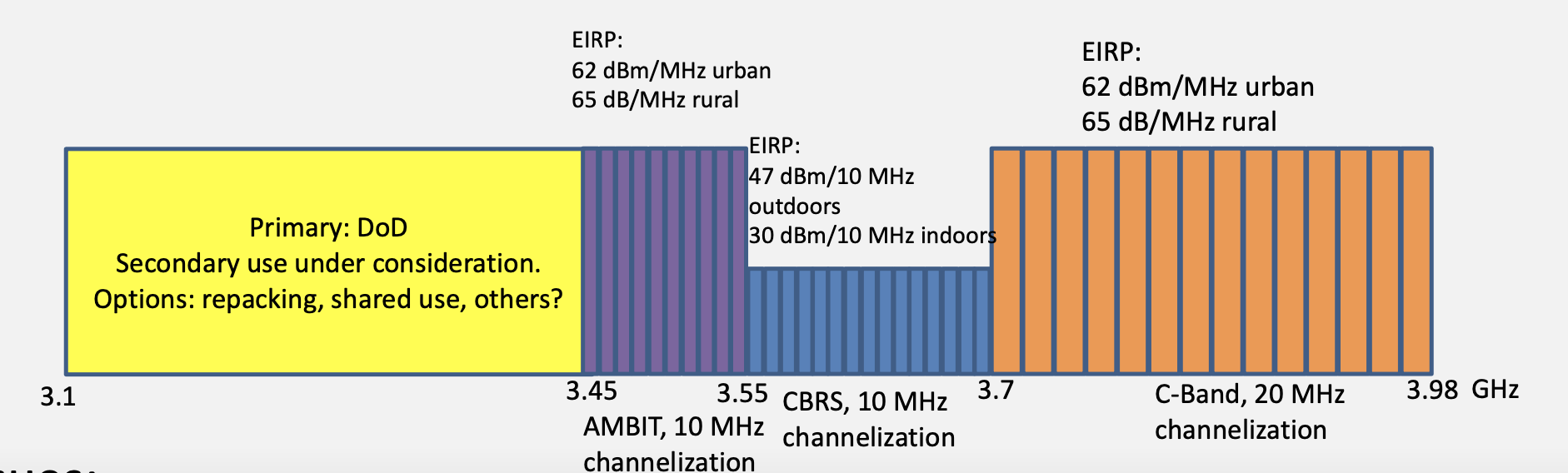}
\caption{Spectrum Chart from 3.1 to 3.9 GHz}
\label{fig:spectrum}
\vspace{-2em}
\end{figure}

The CBRS band was allocated prior to either the C-band or 3.45 GHz. Since the primary incumbent in the CBRS band is Navy radar, a 3-tier access mechanism was put in place to ensure shared access to the band as follows: the highest priority, or Tier 1 users are the Navy radars, followed by Tier 2 Priority Access Licensees (PAL) who acquired licenses through an auction process and finally Tier 3 General Authorized Access (GAA) users who are allowed to access channels that are not being used by Tier 1 or Tier 2 users. This access mechanism is orchestrated by the Spectrum Access System (SAS) which ensures that higher priority users do not face interference from lower priority users. All 150 MHz between 3.55 - 3.7 GHz can be used by Tier 3 GAA users but must not interfere with Tier 1 and Tier 2 as directed by the SAS. To protect the incumbent from harmful interference, the transmit power level of Tiers 2 and 3 is limited to 30 dBm/10 MHz indoors and 47 dBm/10 MHz outdoors, which is considerably lower than that allowed in the adjacent C-band and 3.45 GHz as shown in Fig.~\ref{fig:spectrum}.

Prior to reallocation for terrestrial mobile services, the C-band was primarily used for satellite communications, which was not densely deployed and did not pose an adjacent channel interference concern for CBRS. Similarly, the 3.45 GHz band was a federal band used sparsely in most areas of the country. However, the new rules and auctions permit 5G cellular services to be deployed in both of these bands at much higher power levels of 62 dBm/MHz for urban and 65 dBm/MHz for rural areas as shown in Fig.~\ref{fig:spectrum}. The lack of guard bands between these frequency allocations, combined with the power differences, leads to the potential of adjacent channel interference between CBRS and both upper and lower adjacent bands especially since Time Division Duplex (TDD) systems are being deployed in these bands. 
C-band services are widely deployed in many areas in the US, while 3.45 GHz services are just beginning.

\begin{figure*}[t]
    \centering
    \begin{subfigure}{.63\textwidth}
    \includegraphics[width=.9\linewidth]{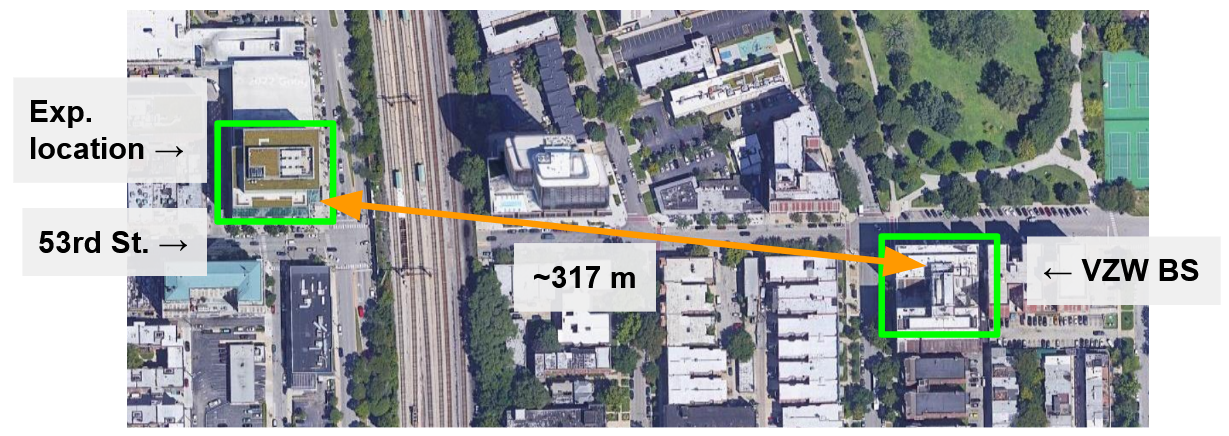}
    \caption{Map of C-band deployment and experiment location.}
    \label{fig:harperMap}
    \end{subfigure}
    \hfill
    \begin{subfigure}{.35\linewidth}
    \includegraphics[width=.9\linewidth]{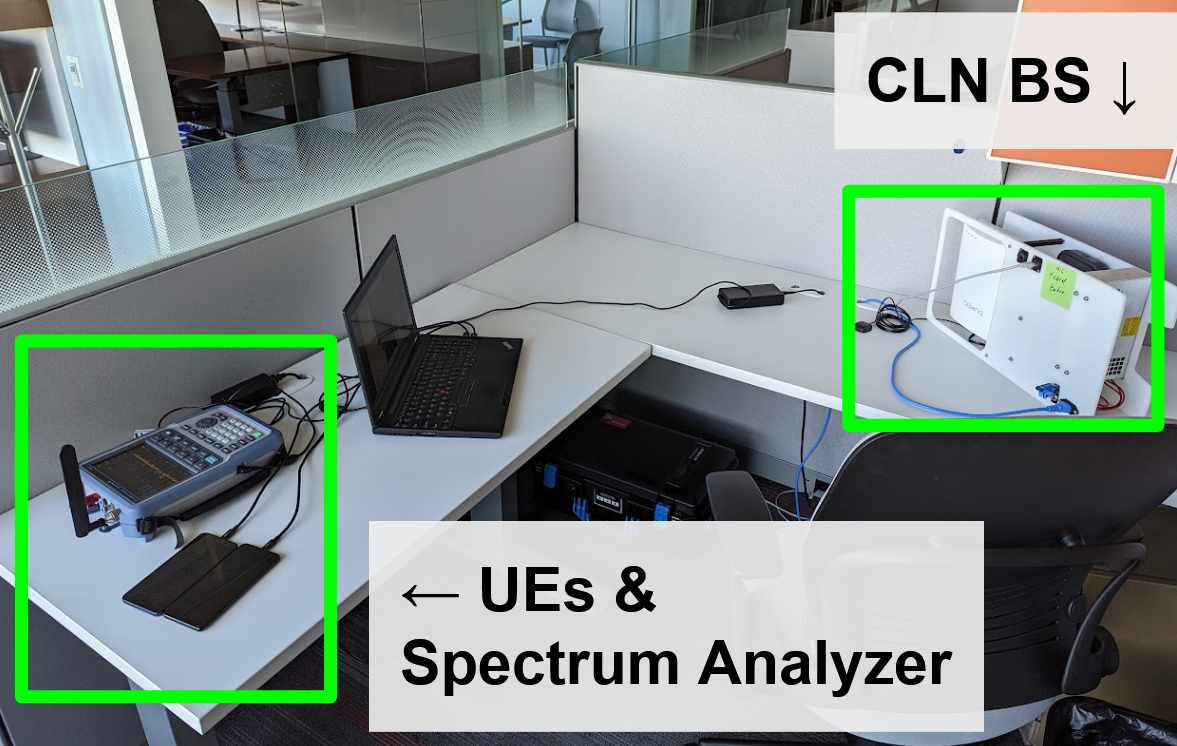}
    \caption{Experiment setup: BS and UEs.}
    \label{fig:harperSetupUEBS}
    \end{subfigure}
    \caption{Experiment map and setup.}
    \vspace{-1em}
\end{figure*}

\begin{table}[t]
    \centering
    \caption{Experiment parameters}
    \label{tab:expParams}
    \begin{tabular}{|p{.22\linewidth}|*{2}{p{.28\linewidth}|}}
        \hline
        \textbf{Parameter} & \multicolumn{2}{c|}{\textbf{Value}} \\
        \hline \hline
        Operators & \vzw & \cln \\
        \hline
        Operating band & C-band & CBRS GAA \\
        \hline
        Radio tech. & 5G & 4G \\
        \hline
        Center freq. & 3.73 GHz & 3.69 and 3.67 GHz \\
        \hline
        Bandwidth & 60 MHz & 20 MHz \\
        \hline
        TDD config. scenarios & 7.4 DL + 2.2 UL & Sa1: 4 DL + 4 UL, Sa2: 6 DL + 2 UL \\
        \hline
        TDD periodicity & 5 ms & 10 ms \\
        \hline
        BS deployment & Outdoor & Indoor \\
        \hline
        Max. BS Power & 79 dBm$^a$ & 23 dBm \\
        \hline
        UE & Samsung S22+ & Samsung S22+ \\
        \hline
        Traffics scenarios & \dl, \ul & \ping, \ping + \dl, \ping + \ul \\
        \hline
        \ping target & N/A & \cln edge server \\
        \hline
        \dl/\ul server & iperf01.uchicago.edu & \cln edge server \\
        \hline
        \dl/\ul parameters & \multicolumn{2}{p{.6\linewidth}|}{target bandwidth 2 Gbps, TCP buffer size 8196 bytes, 10 parallel conns., 500 packets burst} \\
        \hline
        UE location scenarios & \multicolumn{2}{p{.6\linewidth}|}{\vzw @ A, \cln @ A; \vzw @ B, \cln @ B; \vzw @ B, \cln @ A, see Fig.~\ref{fig:harperSetupDiagram}} \\
        \hline
        Exp. run time & \multicolumn{2}{p{.6\linewidth}|}{10 minutes  per combination of scenarios} \\
        \hline
        Total exp. time & \multicolumn{2}{p{.6\linewidth}|}{660 minutes} \\
        \hline
        Time of exp. & \multicolumn{2}{p{.6\linewidth}|}{Between 1 am - 6 am} \\
        \hline
    \end{tabular}\\
    \footnotesize{$^a$We assume the maximum TX power based on the 62 dBm/MHz limit.}
    \vspace{-1.5em}
\end{table}

In this paper, we leverage a real-world C-band deployment to perform experiments that allow an in-depth understanding of the potential impact of adjacent channel interference between CBRS and C-band: a similar situation will be present between 3.45 GHz and CBRS as well, once 3.45 GHz becomes widely deployed. We identified an University of Chicago (UChicago) building with line-of-sight (LOS) to a macro-cell deployment of a C-band base-station (BS). An indoor CBRS device (CBSD, or CBRS base-station), was deployed in an area with LOS to the C-band BS. A spectrum analyzer (SA) was used to first quantify the adjacent power leakage between C-band and CBRS. Detailed experiments were then run with consumer smartphones to connect to both services and quantify the effect of different CBRS TDD configurations and operating channel on uplink (UL) throughput, downlink (DL) throughput and latency. Finally, we demonstrate the improvements to latency when subject to adjacent channel interference by implementing end-to-end microslicing

\section{Background and Related work}
\label{sec:related_work}

C-band and CBRS are both TDD systems, i.e., uplink and downlink transmissions occur over the same frequency channel but separated in time. In most prior TDD cellular deployments, a single operator deployed their network over a wide area and hence the TDD Configuration (the partition between uplink and downlink transmissions) is centrally managed so that co-channel and/or adjacent channel interference is minimized: usually this is done by using the same TDD configuration across the entire deployment. This kind of TDD synchronization ensures that all devices are transmitting traffic in only one direction at any time, either uplink or downlink. However, CBRS users do not need to synchronize amongst each other, or with adjacent C-band, and in fact the use cases may {\bf{require}} different TDD configurations: for example, a video-camera surveillance use case will require higher uplink throughput compared to a video-streaming use case that needs higher downlink throughput. If these use-cases are deployed in adjacent channels, then they may mutually interfere. 

The above interference scenario in cellular deployments is not one that has been studied comprehensively in the literature. A few papers discuss similar problems and propose reducing adjacent channel interference by better filtering~\cite{shamsan2008co,almeida2016mitigating}. We believe that the results presented in this paper are the first to show the effect of different TDD configurations on adjacent channel interference in a real world environment.

\section{Deployment Overview, Measurement Tools, and Methodology}
\label{sec:methodology}

We leverage an outdoor C-band BS deployed on top of a 10-storey building at the intersection of 53rd and E Hyde Park Ave in Chicago. In order to study adjacent channel interference between CBRS and this C-band deployment, we deployed a Celona CBSD indoors on the 9th floor of a UChicago building at 5235 S Harper Court, where the C-band transmission can be received indoors with sufficient signal strength. The set-up is shown in  Figs.~\ref{fig:harperMap} and \ref{fig:harperSetupUEBS}. The CBSD is deployed in a cubicle facing the window with LOS to C-band.
Table~\ref{tab:expParams} summarizes the parameters of both systems.
The two operators are labeled as: \vzw (\textbf{Verizon}), with the C-band deployment, and \cln, the private CBRS network with its CBSD/BS connected to the University of Chicago backhaul.

\begin{figure}[ht]
    \centering
    \includegraphics[width=.8\linewidth]{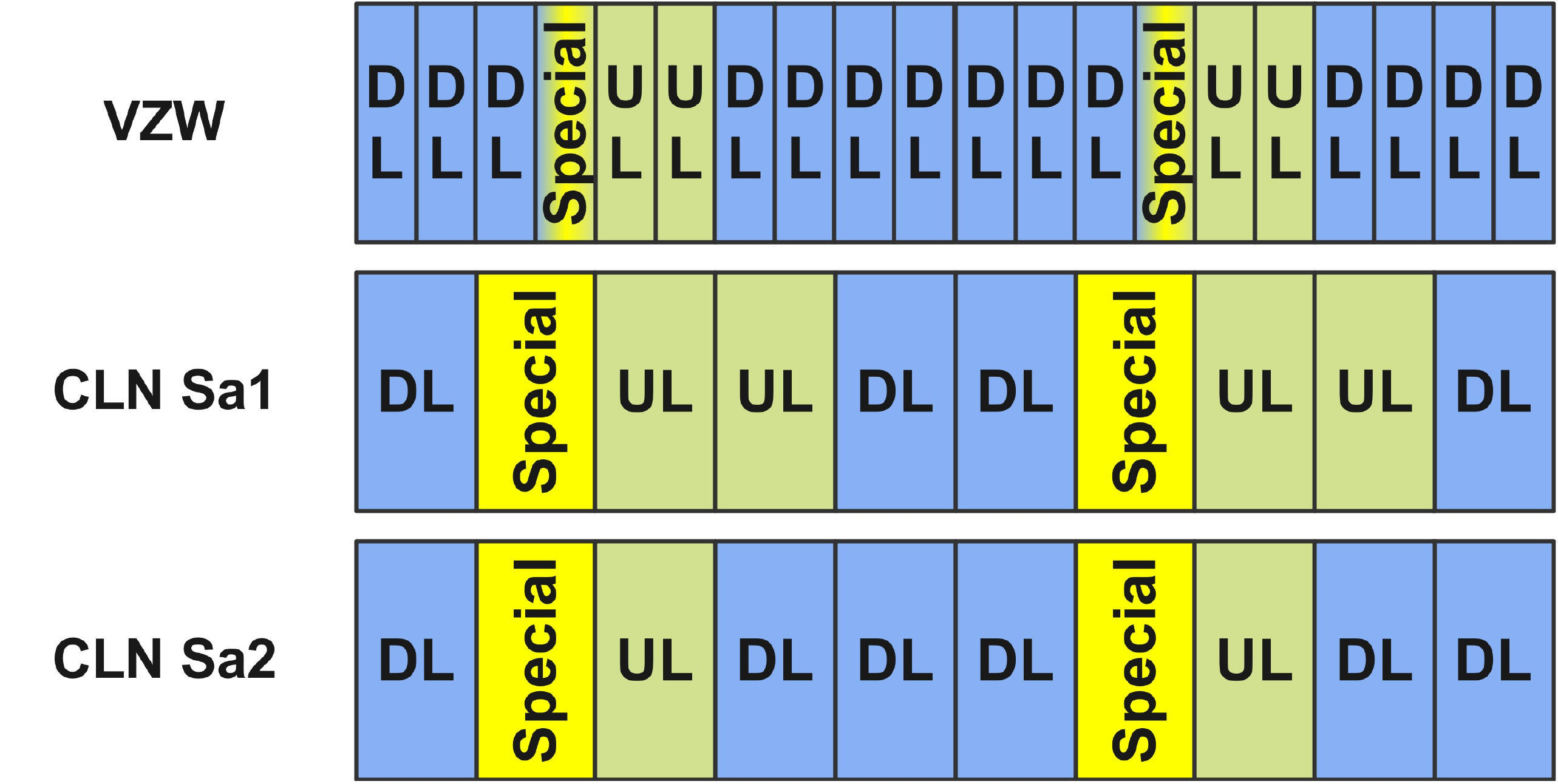}
    \caption{Comparison of TDD configuration.}
    \label{fig:tddComparison}
    \vspace{-1em}
\end{figure}

\noindent{\textit{Overview of Deployments}:}
The \vzw deployment is a 5G NR non-standalone (NSA) configuration with a primary LTE Frequency Division Duplex (FDD) channel in band 66 (DL: 2.11 - 2.13 GHz, UL: 1.71 - 1.73 GHz), and a secondary NR TDD channel in band n77/C-band (3.7 - 3.76 GHz) with 30 kHz sub-carrier spacing. In our throughput analysis, we only consider data transmitted over the 60 MHz C-band and omit the LTE data. The TDD configuration used by \vzw in C-band is shown in Fig.~\ref{fig:tddComparison}: 7 slots for DL and 2 slots for UL, with a slot length of 0.5 ms. Additionally, the "Special" slot is defined to allow greater freedom for resource allocation: 6 symbols are reserved for DL, 4 symbols for UL, and 4 symbols for messaging. In total, there are 7.4 slots reserved for DL and 2.2 slots reserved for UL.

To evaluate potential adjacent channel interference, the \cln CBRS was deployed on the immediately lower adjacent channel, 3.68 - 3.7 GHz using the General Availability Access (GAA) tier of CBRS. As a comparison, we also deployed it on 3.66 - 3.68 GHz, essentially adding a 20 MHz guard band between the CBRS and C-band channels. Since the CBSD was under our control, we varied the TDD configuration of the CBSD between \textbf{Sa1} and \textbf{Sa2} where \textbf{Sa1} uses 4 DL and 4 UL subframes, and \textbf{Sa2} uses 6 DL and 2 UL subframes per radio frame with a subframe length of 1 ms, as shown in Fig.~\ref{fig:tddComparison}. Additionally, while both 5G slot and 4G subframes have the same 0.5 ms duration, issues may arise with the lack of synchronization between \cln and \vzw. Since the CBSD was 4G, the TDD configurations could not be exactly matched to 5G in C-band.

\begin{figure}[t]
    \centering
    \includegraphics[width=.6\linewidth, height = 7cm]{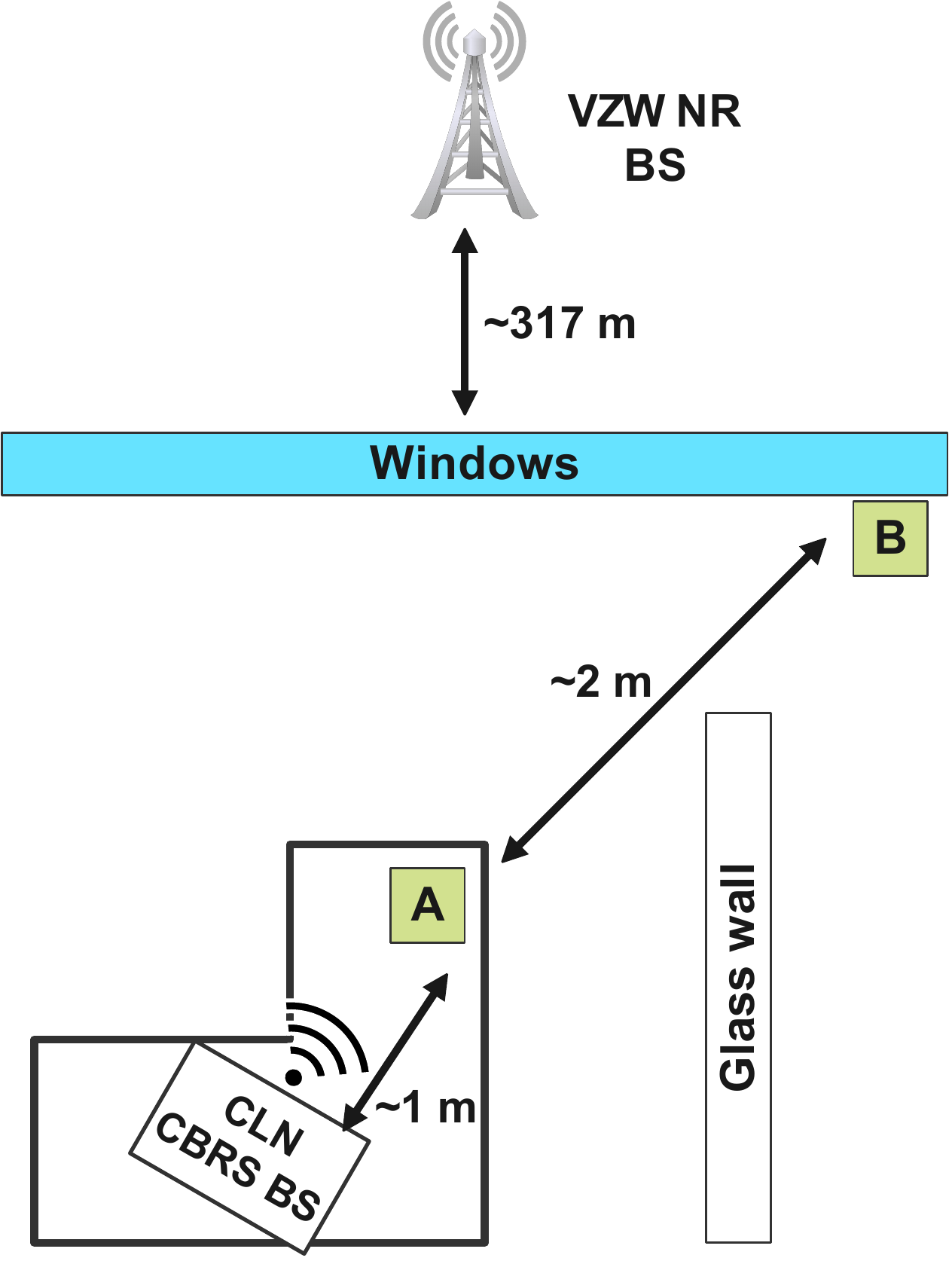}
    \caption{Experiment set-up.}
    \label{fig:harperSetupDiagram}
    \vspace{-1em}
\end{figure}

\noindent{\textit{\bf Measurement Tools and Methodology}:}
Two Samsung S22+ phones (running Android 12) are used as user equipment (UEs), one with a \cln SIM and the other with a \vzw SIM. Both SIMs have unlimited data plans with no throttling of data rates.
We also use a spectrum analyzer (R\&S Spectrum Rider FPH) to measure power over the CBRS and C-band channels.
Fig.~\ref{fig:harperSetupDiagram} is a schematic of the deployment scenario. The CBSD is placed on top of a desk in the cubicle and UEs are deployed in two locations, A and B. Location A is $\sim$1 m from the CBSD, while location B is on top of a desk in an office $\sim$3 m from the CBSD. The spectrum analyzer is always at location A.
Both locations are LOS to the \vzw BS.
We define three measurement scenarios: (1) both UEs at A representing the best condition for \cln UE, (2) both UEs at B representing the best condition for \vzw UE, and (3) \cln UE at A and \vzw UE at B representing the best condition for both UEs to their respective BSs.

Signal measurements are obtained from the Android phones using a commercial measurement app called Qualipoc~\cite{qualipoc} which utilizes the UE's root privilege to establish a low-level interface with the \textit{Qualcomm Diag} utility thus enabling extraction of detailed signal parameters such as primary and secondary channel's RSRP, RSRQ, SINR, MCS, Resource Block (RB) allocation, block error rate (BLER), TDD Config, and physical layer throughput. The DL and UL throughput values mentioned in this paper are physical layer throughput values extracted from Qualipoc. Qualipoc is also capable of actively creating traffic using iperf~\cite{iperf} and ping tools.

Experiments were run for 10 minutes per scenario, with a total experiment time of 660 minutes. The experiments were conducted between 1~am and 6~am to reduce the impact on performance due to the presence of other \vzw users.
Two data transmission scenarios were defined using iperf: \dl and \ul which generate full-buffer downlink (iperf server to UE) and uplink (UE to iperf server) transmission, respectively, with parameters defined in Table~\ref{tab:expParams}.
Table~\ref{tab:expParams} also defines different iperf target servers for each operator, since there is a need to separate the backhaul used for each operator: the \vzw UE uses the UChicago iperf server (\textit{iperf01.uchicago.edu}), while the \cln UE uses an edge server as its target server. The edge server is connected directly to the \cln BS, so the \cln throughput closely reflects the wireless link performance, while \vzw throughput includes the wireless + backhaul performance.
Due to this difference, we do not compare the performance between the operators, rather we compare the relative performance for each operator between the two cases: (i) ``single'' case where one operator is active while the other is idle, and (ii) ``coexistence'' case where both operators are active concurrently.

Additionally, we also measure the performance of \cln for latency-sensitive applications using \ping traffic (64 kbyte ping packets every 10 ms over 10 minutes per scenario) to a separate edge server. The latency metric collected by the \ping tool is defined as the round trip time between UE and the \ping target. To further emulate an intensive low-latency application, we implemented MicroSlicing~\cite{celonaMicroslicing}, a network slicing technology that allows precise control over end-to-end resource and service allocation based on specific Quality of Service (QoS) metrics for different applications and devices. Network administrators can use the Celona Orchestrator or the developer APIs to customize network settings on a device or application specific basis. The orchestrator offers control and adjustments for numerous service types, including data throughput, quality, latency, reliability, and network access policies among others. This enables users to set aside guaranteed portions of the network dedicated to the smooth functioning of the respective device and application. 
The platform also records application-specific service level agreements (SLA) and key performance indicators (KPI) across all devices, granting complete user visibility of device performance across the spectrum. In this experiment, MicroSlicing based resource allocation policy is specified to prioritize \ping over \dl and \ul traffic.



\begin{figure}[t]
    \centering
    \begin{subfigure}{.49\linewidth}
    \includegraphics[width=\linewidth]{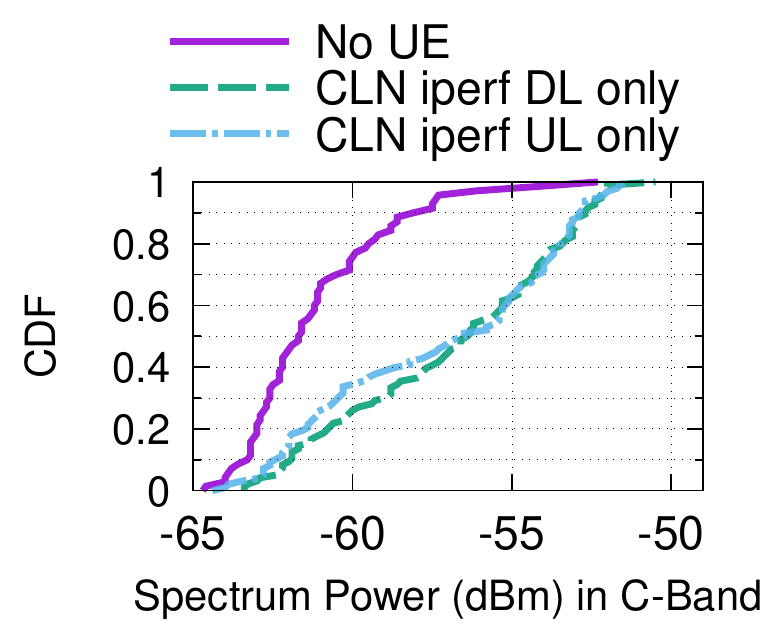}
    \caption{C-band power due to leakage from \cln.}
    \label{fig:spectrumCBand}
    \end{subfigure}
    \hfill
    \begin{subfigure}{.49\linewidth}
    \includegraphics[width=\linewidth]{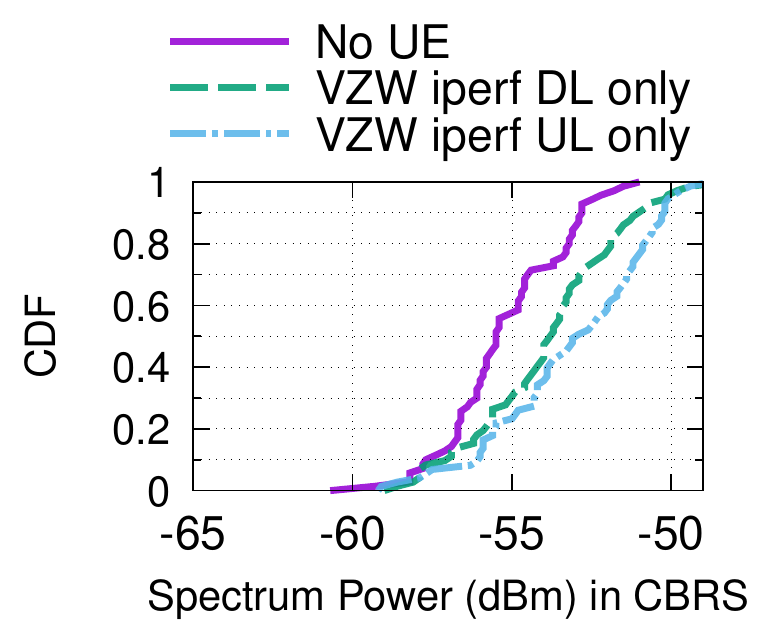}
    \caption{CBRS power due to leakage from \vzw.}
    \label{fig:spectrumCBRS}
    \end{subfigure}
    \caption{Spectrum Analyzer measurements of mutual OOB leakage between CBRS and C-band.}
    \label{fig:spectrumAdjacent}
    \vspace{-1.5em}
\end{figure}

\begin{figure*}[t]
    \centering
    \begin{subfigure}{.245\textwidth}
    \includegraphics[width=\linewidth]{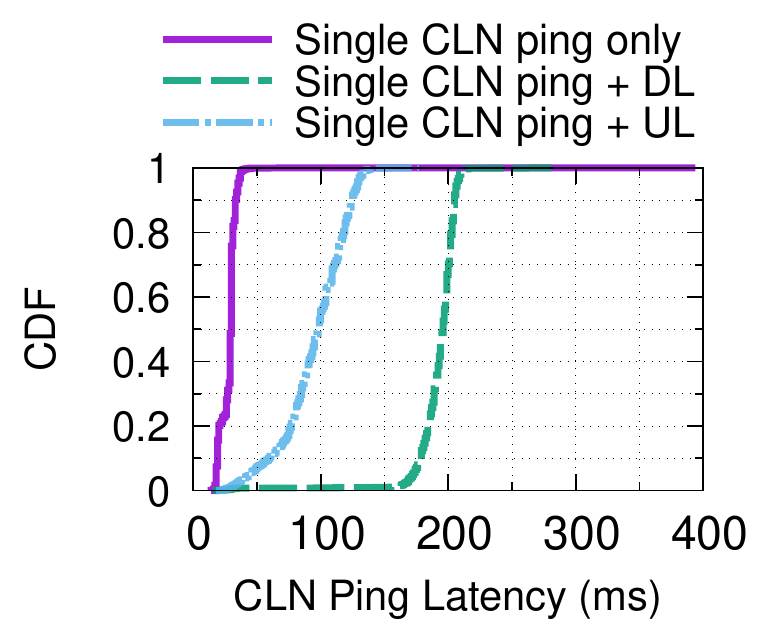}
    \caption{Single cases with varying traffics, no MicroSlicing}
    \label{fig:pingSingle}
    \end{subfigure}
    \hfill
    \begin{subfigure}{.245\textwidth}
    \includegraphics[width=\linewidth]{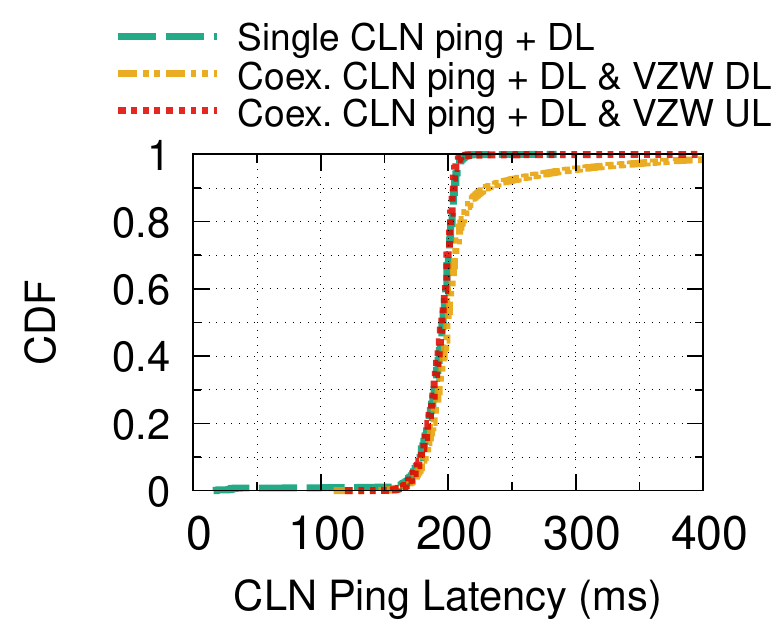}
    \caption{Single \vs Coex. cases on \ping + \dl, no MicroSlicing}
    \label{fig:pingCoexDl}
    \end{subfigure}
    \hfill
    \begin{subfigure}{.245\textwidth}
    \includegraphics[width=\linewidth]{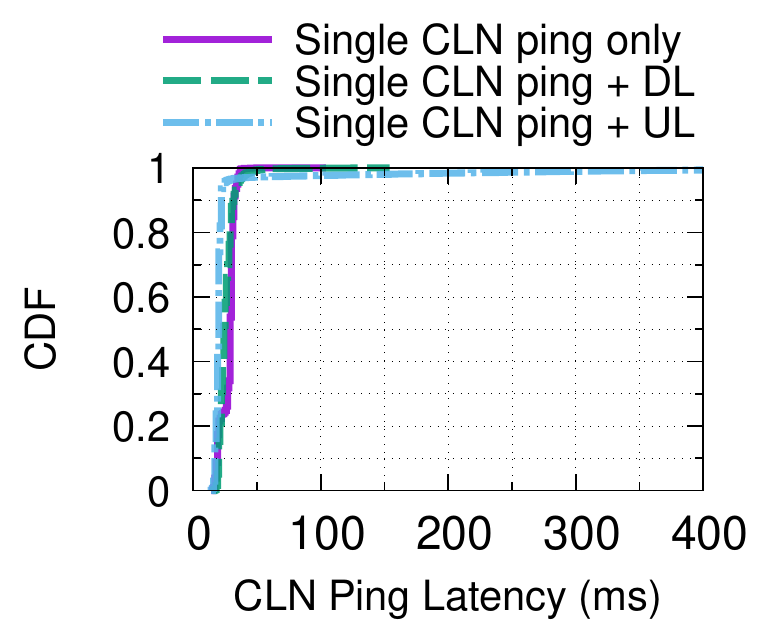}
    \caption{Single cases with varying traffics and MicroSlicing}
    \label{fig:pingSingleMS}
    \end{subfigure}
    \hfill
    \begin{subfigure}{.245\textwidth}
    \includegraphics[width=\linewidth]{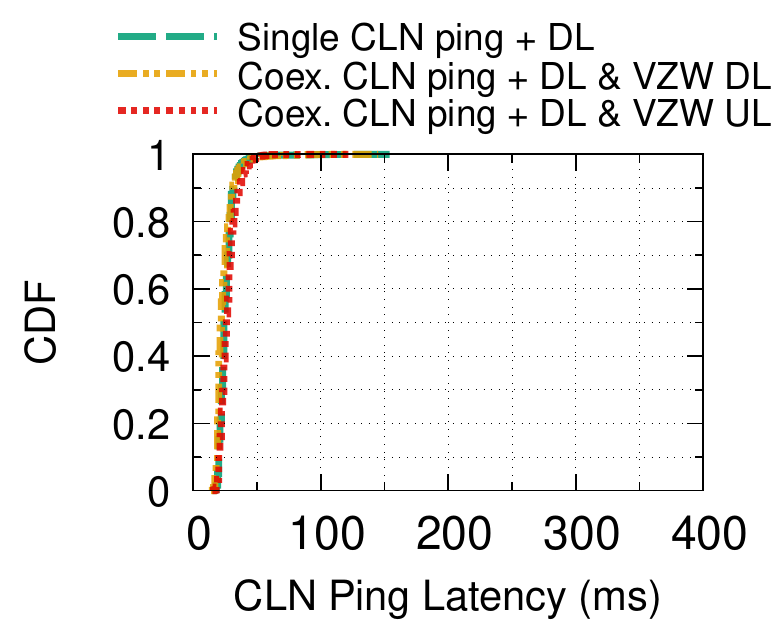}
    \caption{Single \vs Coex. cases on \ping + \dl, with MicroSlicing}
    \label{fig:pingCoexDlMS}
    \end{subfigure}
    \caption{Ping latency performance of \cln.}
    \label{fig:pingPerf}
    \vspace{-1em}
\end{figure*}

\section{Experimental Results}\label{sec:results}


\subsection{Out-of-band (OOB) interference quantified by spectrum analyzer measurements}

Fig.~\ref{fig:spectrumAdjacent} shows the spectrum analyzer measurements of OOB interference due to transmissions to and from the UEs. These spectrum measurements are done with all UEs and the spectrum analyzer in close proximity to each other in location A. The spectrum analyzer measures the power on the 3.68-3.7 GHz CBRS channel and the 3.7-3.76 GHz C-band channel in the following scenarios: (i) both UEs are turned off, (ii) only the \cln UE transmits \dl / \ul, and (iii) only the \vzw UE transmits \dl / \ul. We also vary \cln TDD configurations between Sa1 and Sa2.
Fig.~\ref{fig:spectrumCBand} shows the OOB effect on C-band channel due to \cln transmission on the CBRS channel using both TDD configurations, \ie there is clearly an increase of power observed in the adjacent C-band channel compared to when both UEs are turned off. Similarly, Fig.~\ref{fig:spectrumCBRS} shows the effect of \vzw transmission in C-band on the CBRS channel, again demonstrating a power increase.
This initial power analysis clearly demonstrates the potential of OOB interference on both operators. In the following sub-sections, we demonstrate the impact of increased OOB interference to ping latency and DL throughput performance at the UEs.

\subsection{Latency performance of \cln}

Latency performance of CBRS with and without adjacent channel transmissions was evaluated only in location A, with \cln TDD configuration set to Sa1.
Fig.~\ref{fig:pingSingle} shows the ``single'' case, \ie no interference from the \vzw UE, without MicroSlicing and Fig.~\ref{fig:pingSingleMS} shows the performance in the same scenario but with MicroSlicing. Similarly, Fig.~\ref{fig:pingCoexDl} shows the ``coex'' case, \ie  interference from the \vzw UE, without MicroSlicing and Fig.~\ref{fig:pingCoexDlMS} shows the performance in the same scenario but with MicroSlicing. In both cases, overall, we observe less difference in latency when \ping traffic is transmitted along with \dl and \ul when MicroSlicing is used compared to no MicroSlicing. In particular, we observe increased latency on 20\% of the data, when \vzw is using \dl traffic without MicroSlicing while there is no impact of OOB interference to \cln's latency performance when it is using MicroSlicing (Fig.~\ref{fig:pingCoexDlMS}).
The reduction of the effect of the OOB interference can be explained by the MicroSlicing policy, which assigns a higher priority to \ping packets thus ensuring timely packet arrival, even under OOB interference.

\subsection{Impact of \vzw's OOB interference on \cln's throughput}

\begin{figure*}[t]
    \begin{subfigure}{.245\textwidth}
    \includegraphics[width=\linewidth]{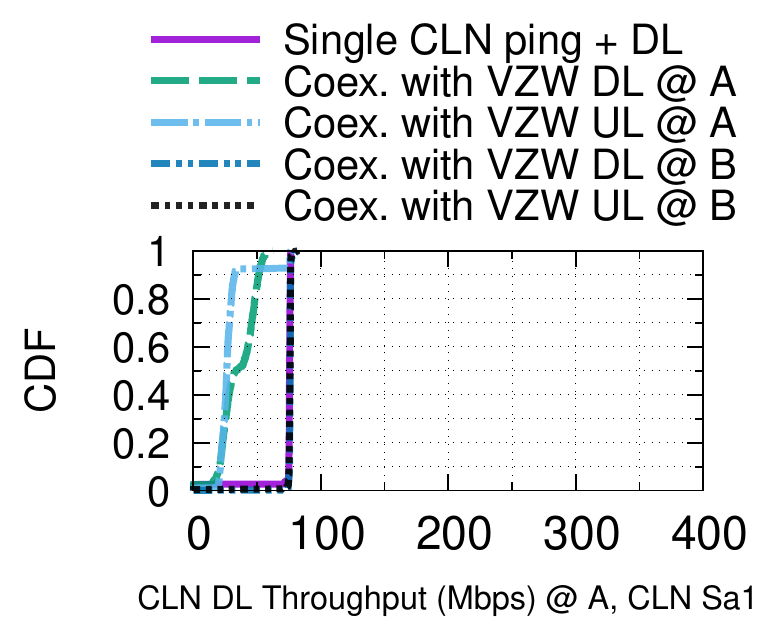}
    \caption{\cln DL tput. @ A, \cln Sa1}
    \label{fig:tputClnDlASa1}
    \end{subfigure}
    \hfill
    \begin{subfigure}{.245\textwidth}
    \includegraphics[width=\linewidth]{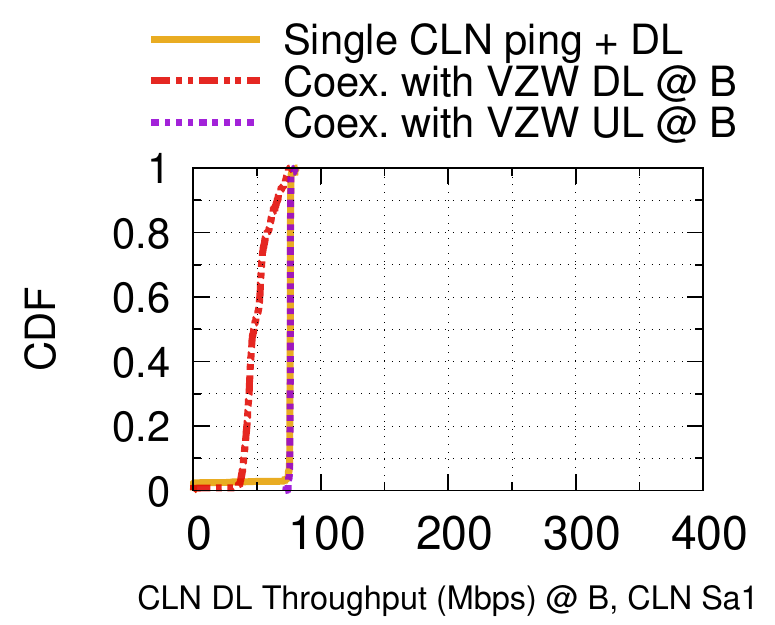}
    \caption{\cln DL tput. @ B, \cln Sa1}
    \label{fig:tputClnDlBSa1}
    \end{subfigure}
    \hfill
    \begin{subfigure}{.245\textwidth}
    \includegraphics[width=\linewidth]{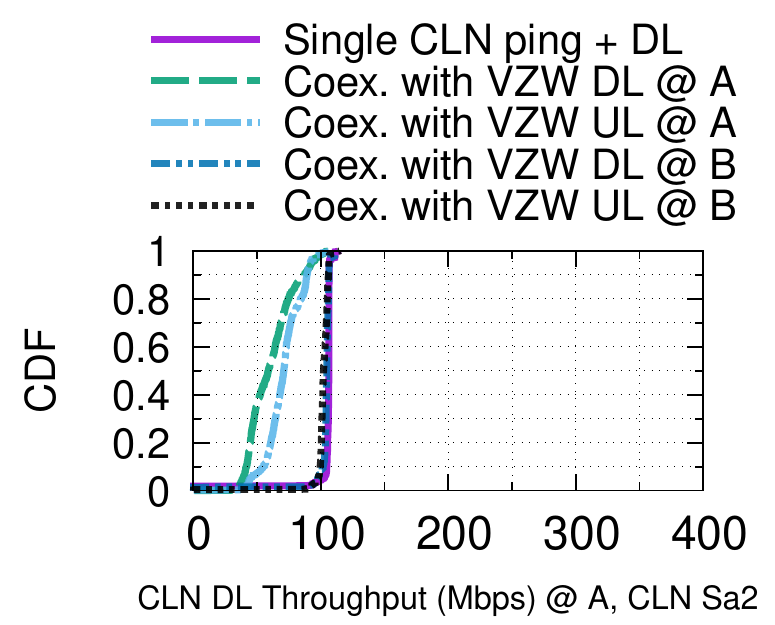}
    \caption{\cln DL tput. @ A, \cln Sa2}
    \label{fig:tputClnDlASa2}
    \end{subfigure}
    \hfill
    \begin{subfigure}{.245\textwidth}
    \includegraphics[width=\linewidth]{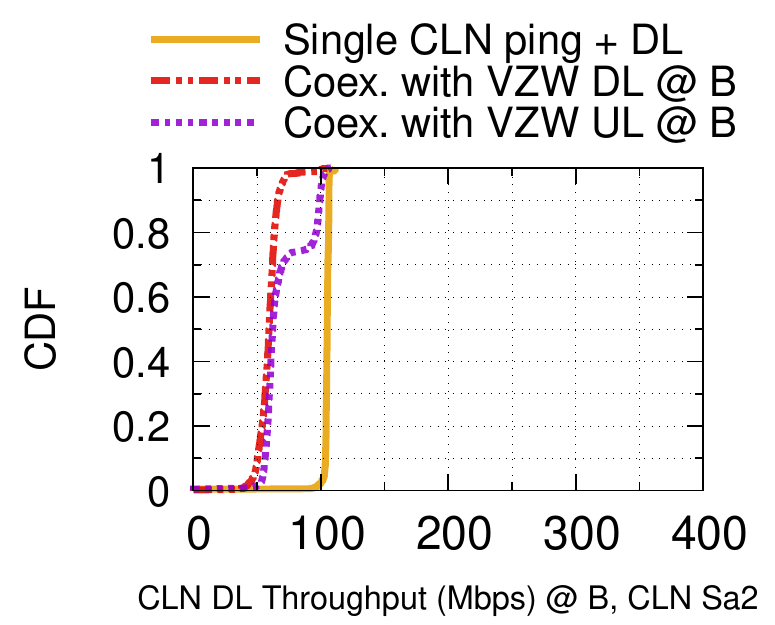}
    \caption{\cln DL tput. @ B, \cln Sa2}
    \label{fig:tputClnDlBSa2}
    \end{subfigure}
    \caption{Coexistence performance in terms of \cln DL throughput under varying \cln TDD configurations.}
    \label{fig:tputClnDlTdd}
    \vspace{-1em}
\end{figure*}


\begin{figure*}[t]
    \centering
    \begin{subfigure}{.325\textwidth}
    \includegraphics[width=\linewidth]{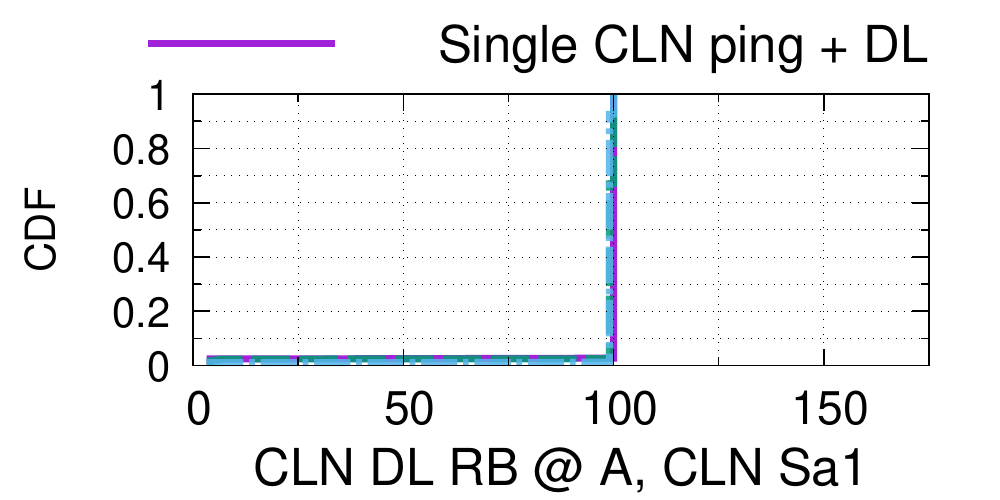}
    \caption{\cln DL RB}
    \label{fig:tputClnDlARbSa1}
    \end{subfigure}
    \begin{subfigure}{.325\textwidth}
    \includegraphics[width=\linewidth]{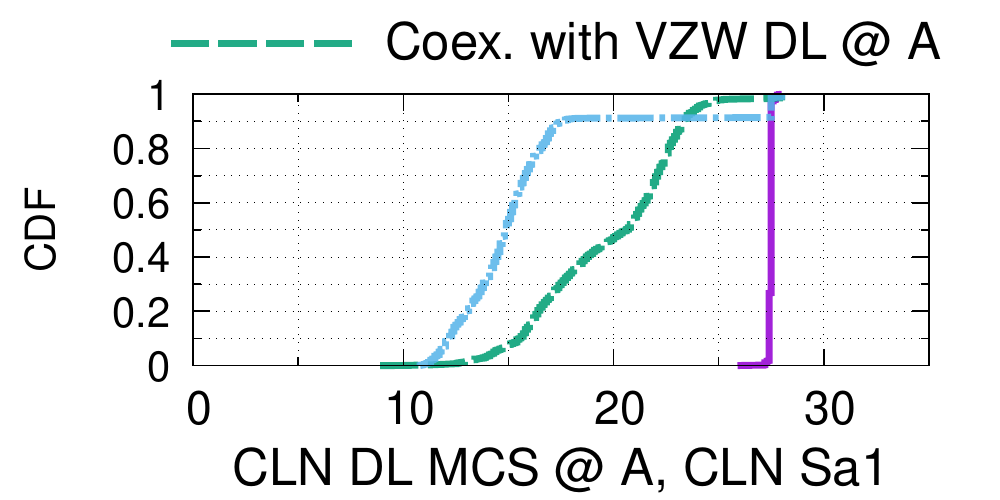}
    \caption{\cln DL MCS}
    \label{fig:tputClnDlAMcsSa1}
    \end{subfigure}
    \begin{subfigure}{.325\textwidth}
    \includegraphics[width=\linewidth]{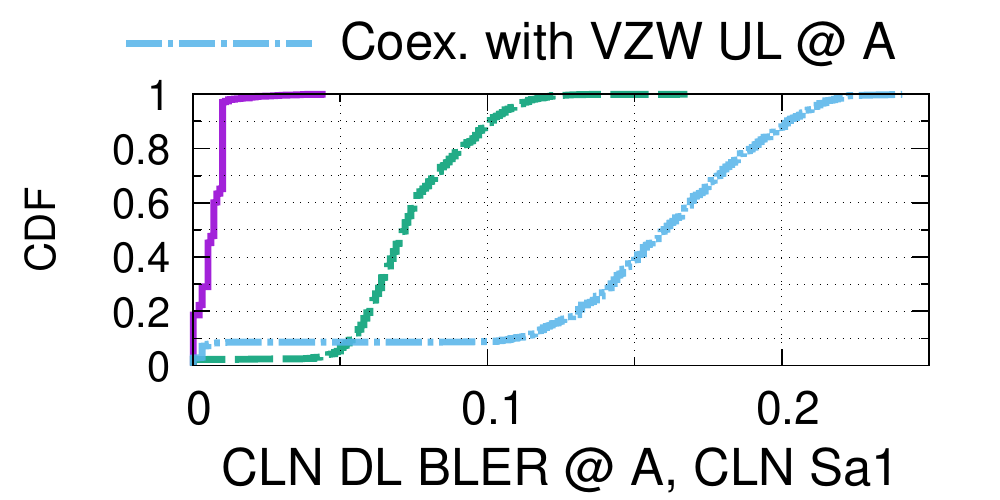}
    \caption{\cln DL BLER}
    \label{fig:tputClnDlABlerSa1}
    \end{subfigure}
    \caption{Representative comparison of DL RB, MCS, and BLER for \cln @ A, on \cln Sa1.}
    \label{fig:tputClnTdd2}
    \vspace{-1em}
\end{figure*}

Fig.~\ref{fig:tputClnDlTdd} shows coexistence performance of \cln in terms of physical layer DL throughput with varying \cln TDD configurations. Due to page limitations, we omit the UL throughput results which did not demonstrate any effect of adjacent channel interference. Only \ping + \dl traffic is analyzed as there is \textbf{no difference} between the \dl only and \ping + \dl, which is expected since the flow of \ping packets are also counted in the throughput metric.
Figs.~\ref{fig:tputClnDlASa1} and \ref{fig:tputClnDlBSa1} show the effect of OOB interference on \cln DL throughput, using Sa1 configuration, at location A and B, respectively. Firstly, we observe no impact of coexistence when \cln UE is at A and \vzw UE is at B, \ie the effect of OOB interference is only observed when the UEs are close to each other.
When \cln is using Sa1, we observe the highest throughput degradation when \cln \ping + \dl @ A and \vzw \ul @ A coexist: 60 \% reduction of \cln DL throughput compared to the "single" case. When the \vzw UE is transmitting \ul @ A, the proximity to the \cln CBSD causes a large throughput reduction which can be explained by the TDD configuration as shown on Fig.~\ref{fig:tddComparison}: in the worst case, \cln's DL capability is reduced to half due to the overlap with two of \vzw's UL slots. This effect is not observed at B due to the greater distance to the CBSD. Only DL traffic from \vzw at B affects \cln which is also at B.
Similarly when \cln is using Sa2, the DL throughput degradation is observed on Figs.~\ref{fig:tputClnDlASa2} and \ref{fig:tputClnDlBSa2}. The greatest throughput degradation in this case is between \cln \ping + \dl @ B and \vzw \dl @ B, which is 43\% reduction from the single case. Additionally, \vzw \ul @ B affects \cln DL throughput on Sa2 (an effect not observed on Sa1), which is possibly due to the higher number of DL subframes in the Sa2 configuration which leads to a higher probability of overlap with \vzw UL slots.


To further demonstrate the effect of OOB interference, we correlate RB, MCS, and BLER values when \cln is using \ping + \dl traffic as shown in Fig.~\ref{fig:tputClnTdd2}. 
We could not correlate OOB interference to the captured RSRP and RSRQ values: while the RSRP and RSRQ are well-defined by 3GPP, we cannot confirm the correctness of its implementation inside the modem. 
A representative coexistence case of \cln \ping + \dl @ A and \vzw \dl @ A on \cln Sa1 is chosen for analysis, but the same conclusion is observed in other cases.
First, Fig.~\ref{fig:tputClnDlARbSa1} shows the full RB allocation to the \cln UE in all cases, since the \cln UE is the only one connected to the CBSD. On the other hand, Figs.~\ref{fig:tputClnDlAMcsSa1} and \ref{fig:tputClnDlABlerSa1} respectively show a degradation of MCS allocation and increase in BLER under coexistence, leading to a reduction in throughput. 
Combined with the spectrum analyzer power analysis and the higher \vzw BS transmit power, we are certain that \cln's DL throughput reduction is caused by OOB interference.

\subsection{Impact of \cln's OOB interference on \vzw throughput}

\begin{figure*}[t]
    \centering
    \begin{subfigure}{.245\textwidth}
    \includegraphics[width=\linewidth]{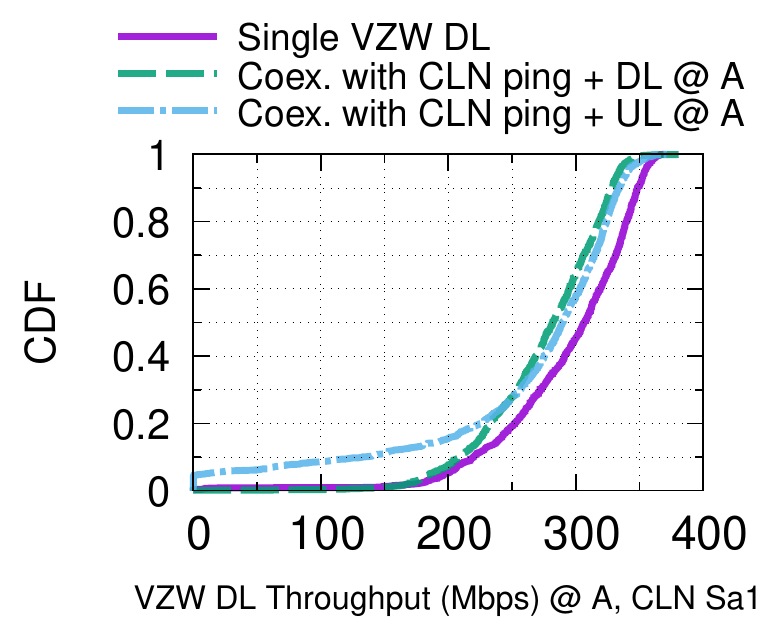}
    \caption{\vzw DL tput. @ A, \cln Sa1}
    \label{fig:tputVzwDlASa1}
    \end{subfigure}
    \hfill
    \begin{subfigure}{.245\textwidth}
    \includegraphics[width=\linewidth]{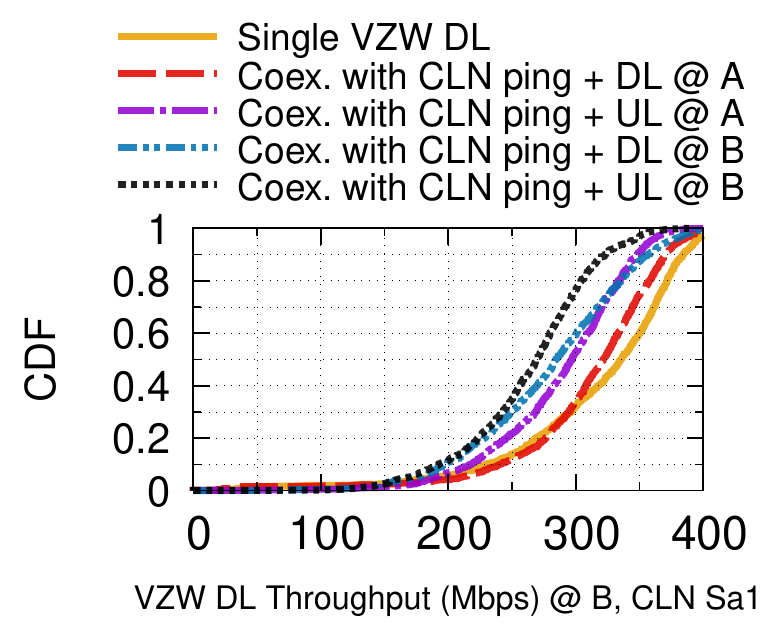}
    \caption{\vzw DL tput. @ B, \cln Sa1}
    \label{fig:tputVzwDlBSa1}
    \end{subfigure}
    \hfill
    \begin{subfigure}{.245\textwidth}
    \includegraphics[width=\linewidth]{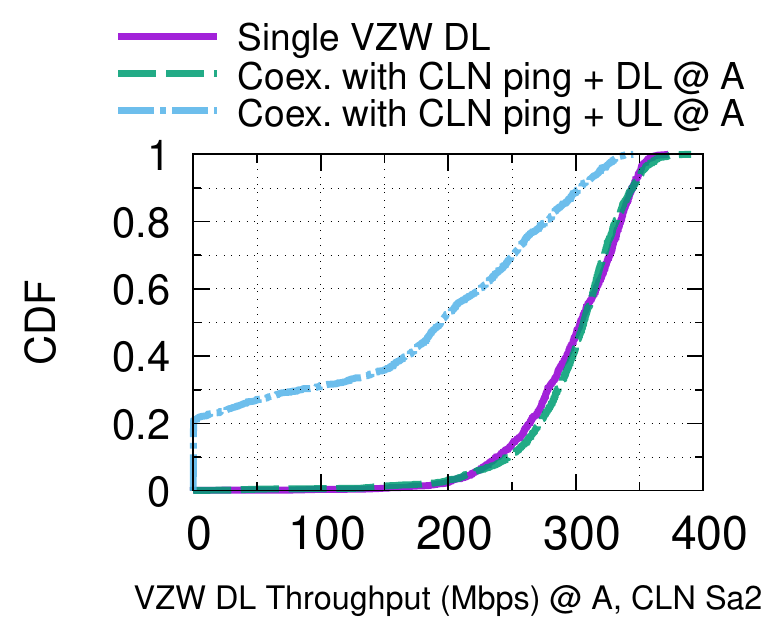}
    \caption{\vzw DL tput. @ A, \cln Sa2}
    \label{fig:tputVzwDlASa2}
    \end{subfigure}
    \hfill
    \begin{subfigure}{.245\textwidth}
    \includegraphics[width=\linewidth]{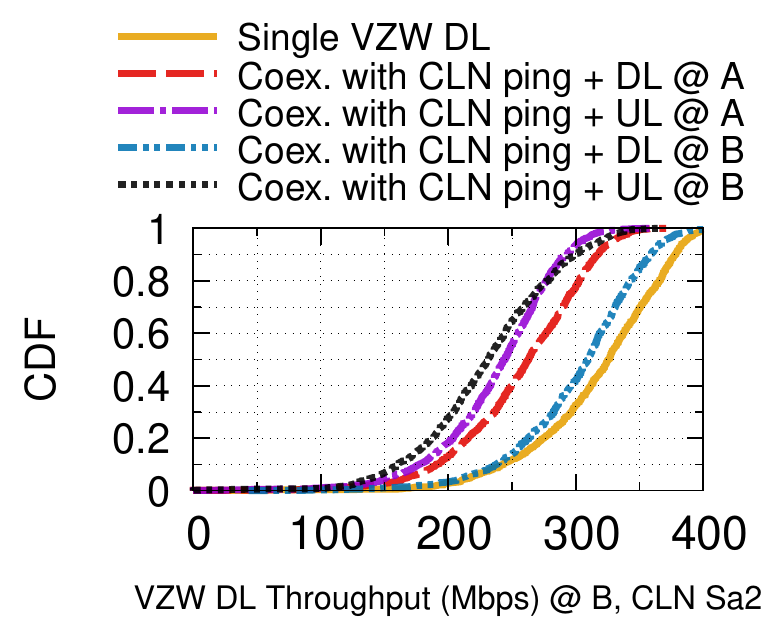}
    \caption{\vzw DL tput. @ B, \cln Sa2}
    \label{fig:tputVzwDlBSa2}
    \end{subfigure}
    \caption{Coexistence performance in terms of \vzw DL throughput under varying \cln TDD configurations.}
    \label{fig:tputVzwDlTdd}
    \vspace{-1em}
\end{figure*}

\begin{figure*}[htb!]
    \centering
    \begin{subfigure}{.325\textwidth}
    \includegraphics[width=\linewidth]{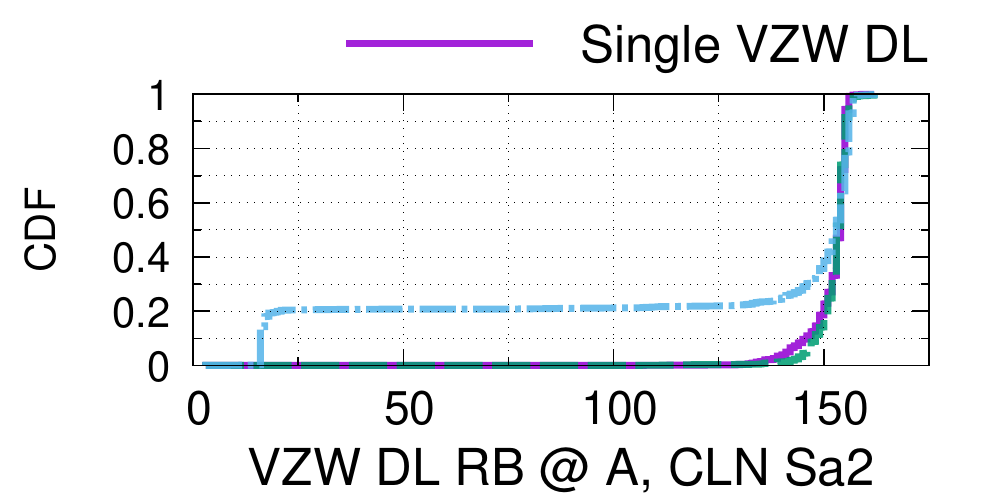}
    \caption{\vzw DL RB}
    \label{fig:tputVzwDlARbSa2}
    \end{subfigure}
    \begin{subfigure}{.325\textwidth}
    \includegraphics[width=\linewidth]{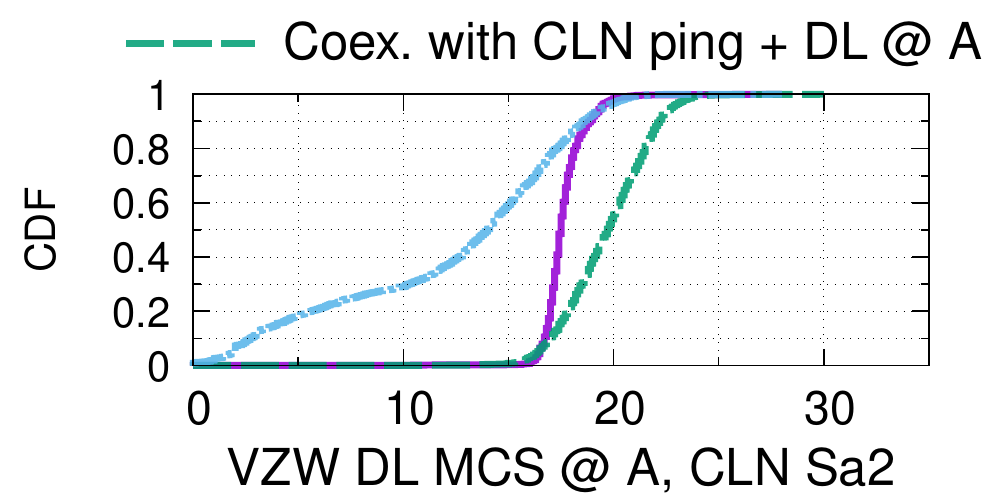}
    \caption{\vzw DL MCS}
    \label{fig:tputVzwDlAMcsSa2}
    \end{subfigure}
    \begin{subfigure}{.325\textwidth}
    \includegraphics[width=\linewidth]{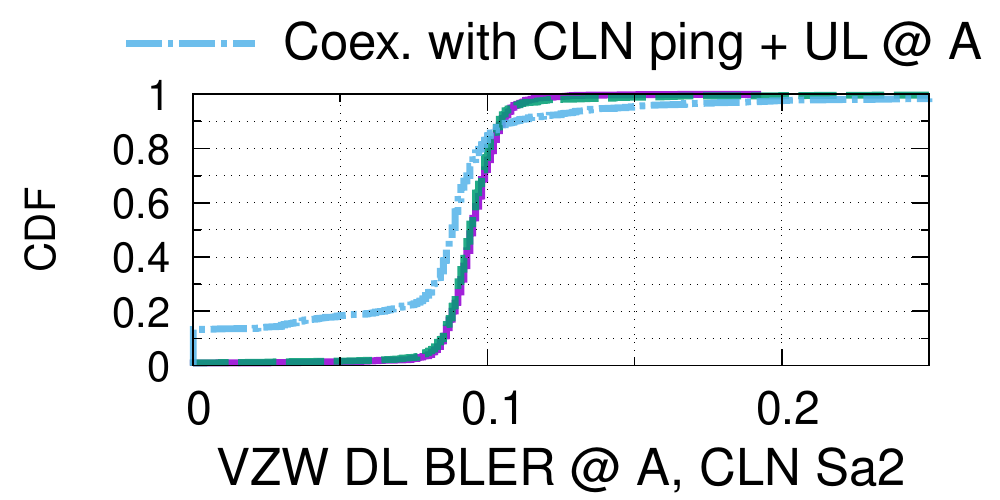}
    \caption{\vzw DL BLER}
    \label{fig:tputVzwDlABlerSa2}
    \end{subfigure}
    \caption{Representative comparison of DL RB, MCS, and BLER for \vzw @ A, on \cln Sa2.}
    \label{fig:tputVzwTdd2}
    \vspace{-1em}
\end{figure*}

\begin{figure*}[t]
    \begin{subfigure}{.245\textwidth}
    \includegraphics[width=\linewidth]{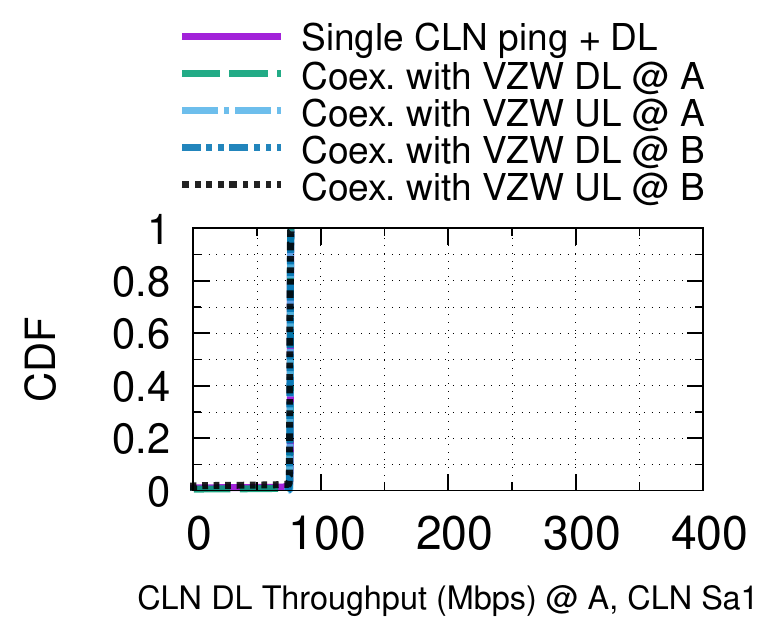}
    \caption{\cln DL tput. @ A, \cln Sa1}
    \label{fig:tputClnDlASa1Gap}
    \end{subfigure}
    \hfill
    \begin{subfigure}{.245\textwidth}
    \includegraphics[width=\linewidth]{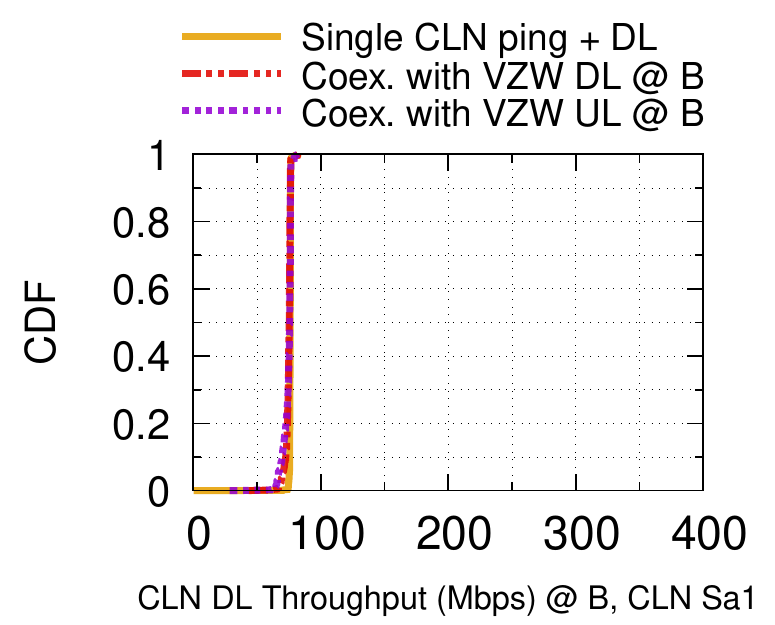}
    \caption{\cln DL tput. @ B, \cln Sa1}
    \label{fig:tputClnDlBSa1Gap}
    \end{subfigure}
    \hfill
    \begin{subfigure}{.245\textwidth}
    \includegraphics[width=\linewidth]{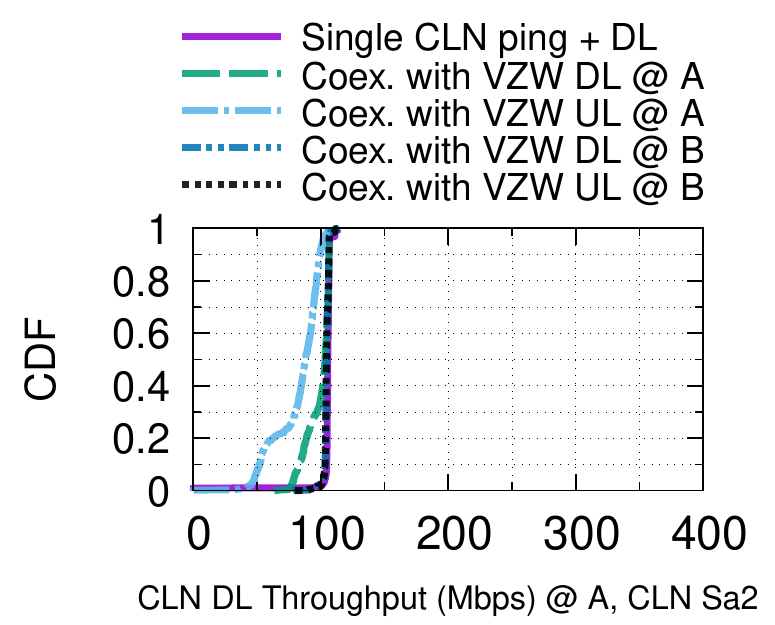}
    \caption{\cln DL tput. @ A, \cln Sa2}
    \label{fig:tputClnDlASa2Gap}
    \end{subfigure}
    \hfill
    \begin{subfigure}{.245\textwidth}
    \includegraphics[width=\linewidth]{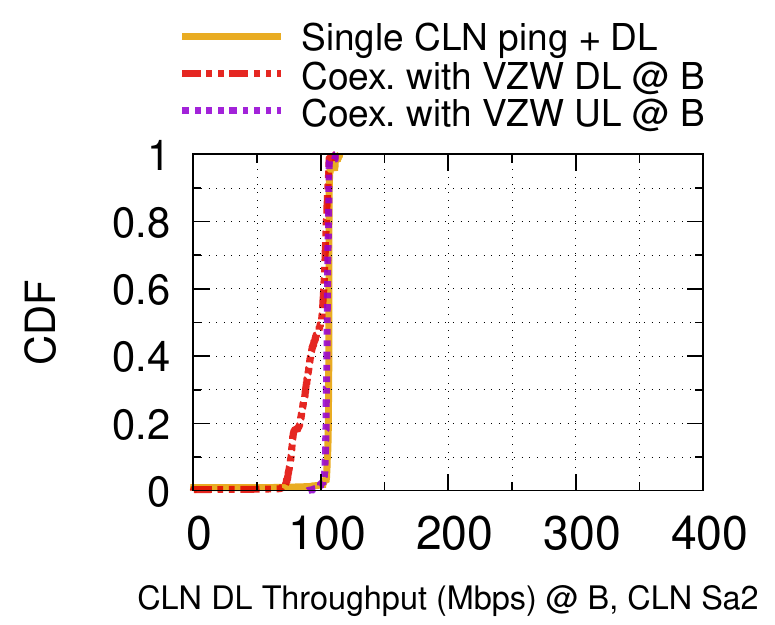}
    \caption{\cln DL tput. @ B, \cln Sa2}
    \label{fig:tputClnDlBSa2Gap}
    \end{subfigure}
    \caption{Coexistence performance in terms of \cln DL throughput with 20 MHz guard band.}
    \label{fig:tputClnDlTddGap}
    \vspace{-1em}
\end{figure*}

\begin{figure*}[t]
    \centering
    \begin{subfigure}{.325\textwidth}
    \includegraphics[width=\linewidth]{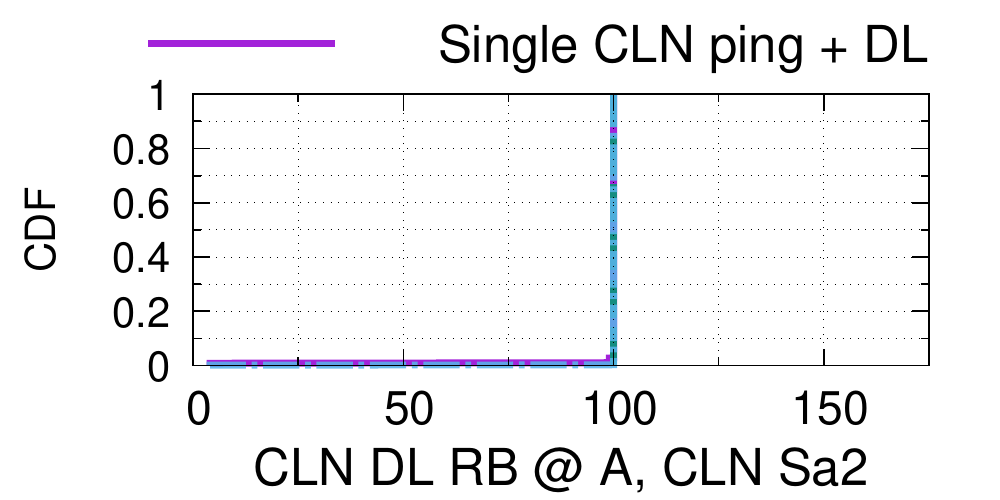}
    \caption{\cln DL RB}
    \label{fig:tputClnDlARbSa2Gap}
    \end{subfigure}
    \begin{subfigure}{.325\textwidth}
    \includegraphics[width=\linewidth]{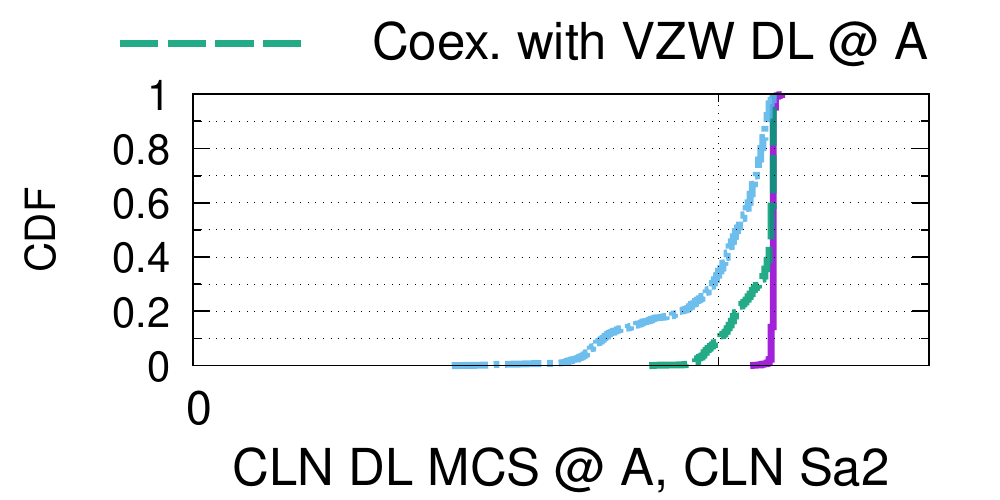}
    \caption{\cln DL MCS}
    \label{fig:tputClnDlAMcsSa2Gap}
    \end{subfigure}
    \begin{subfigure}{.325\textwidth}
    \includegraphics[width=\linewidth]{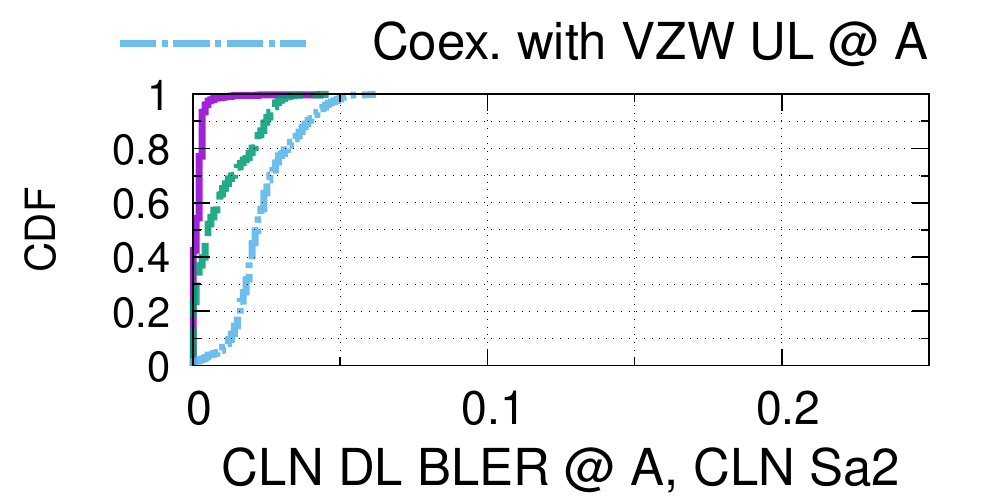}
    \caption{\cln DL BLER}
    \label{fig:tputClnDlABlerSa2Gap}
    \end{subfigure}
    \caption{Representative comparison of DL RB, MCS, and BLER for \cln @ A, on \cln Sa2, with 20 MHz guard band.}
    \label{fig:tputClnTdd2Gap}
    \vspace{-1.5em}
\end{figure*}

Similar to the previous analysis, we only focus on analyzing the coexistence between \vzw \dl and \cln \ping + \dl/\ul (hereinafter shortened to \cln \dl/\ul). We omit the analysis on \vzw \ul due to no impact of OOB interference. Additionally, data containing \cln \ping only traffic is also omitted from our analysis due to no interference from \ping traffic's low network utilization.
Fig.~\ref{fig:tputVzwDlTdd} shows \vzw DL throughput on varying locations and \cln TDD configuration. When \cln is using Sa1 configuration, Fig.~\ref{fig:tputVzwDlASa1} shows a similar reduction of DL throughput when coexisting with \cln \dl and \ul at location A, while Fig.~\ref{fig:tputVzwDlBSa1} shows the largest throughput reduction (in Sa1) of 17\% when coexisting between \vzw \dl @ B and \cln \ul @ B.
We also observe a DL throughput reduction for \vzw in the location scenario \cln @ A and \vzw @ B, although this is lower compared to when both UEs are side-by-side.
For Sa2, the highest DL throughput reduction is observed in scenario \vzw \dl at A and \cln \ul at A as shown on Fig.~\ref{fig:tputVzwDlASa2}, \ie 43\% reduction compared to the single case. Fig.~\ref{fig:tputVzwDlBSa2} shows the counterpart at location B, with the highest reduction of 27\% when coexisting with \cln \ul at B. 

Most of the scenarios described above do not seem to exhibit a drastic change in terms of RB allocation, MCS, and BLER between the coexistence and single cases, except for the scenario of \vzw @ A \& \cln @ A, \cln uses Sa2 configuration. Thus, we focus on analyzing the said scenario as shown in Fig.~\ref{fig:tputVzwTdd2}. 
Fig.~\ref{fig:tputVzwDlARbSa2}, \ref{fig:tputVzwDlAMcsSa2}, and \ref{fig:tputVzwDlABlerSa2} shows the CDF comparison of RB allocation, MCS, and BLER, respectively. There is a slight decrease of RB allocation, MCS, and BLER on coexistence cases compared to the single case.
As we see from Fig.~\ref{fig:tddComparison}, \cln \ul using Sa2 should affects \vzw \dl less than Sa1 due to the fewer number of uplink subframes that overlap with \vzw's downlink slots. However, our experiment is not capable of capturing the exact frame timing to determine interference.
Thus, the effect of OOB interference is not directly apparent here: the \vzw BS may have possibly reacted to the interference by lowering RB allocation and MCS thus resulting in better BLER performance but lower throughput.

\subsection{Impact of OOB interference with a 20 MHz guard band}

In this analysis, we measured the throughput performance of both operators when the \cln operating channel was moved to 3.66 - 3.68 GHz, thus adding a 20 MHz guard band between the CBRS and C-band channels. We refer to these scenarios as \textbf{GAP}, while the prior scenarios as \textbf{non-GAP}.
We omit showing the spectrum analysis in this scenario, since we observed no power leakage in the C-band channel and the new CBRS channel (3.66 - 3.68 GHz) as expected.

\begin{figure*}[t]
    \centering
    \begin{subfigure}{.245\textwidth}
    \includegraphics[width=\linewidth]{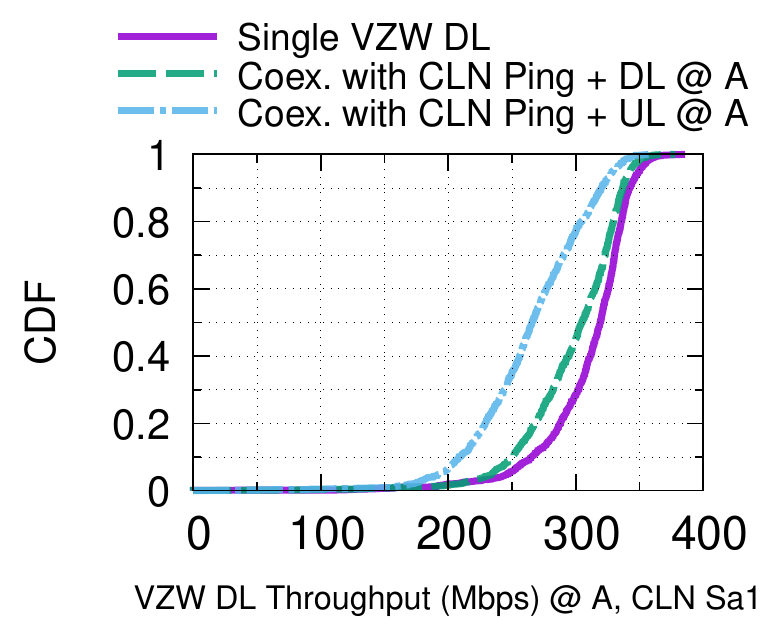}
    \caption{\vzw DL tput. @ A, \cln Sa1}
    \label{fig:tputVzwDlASa1Gap}
    \end{subfigure}
    \hfill
    \begin{subfigure}{.245\textwidth}
    \includegraphics[width=\linewidth]{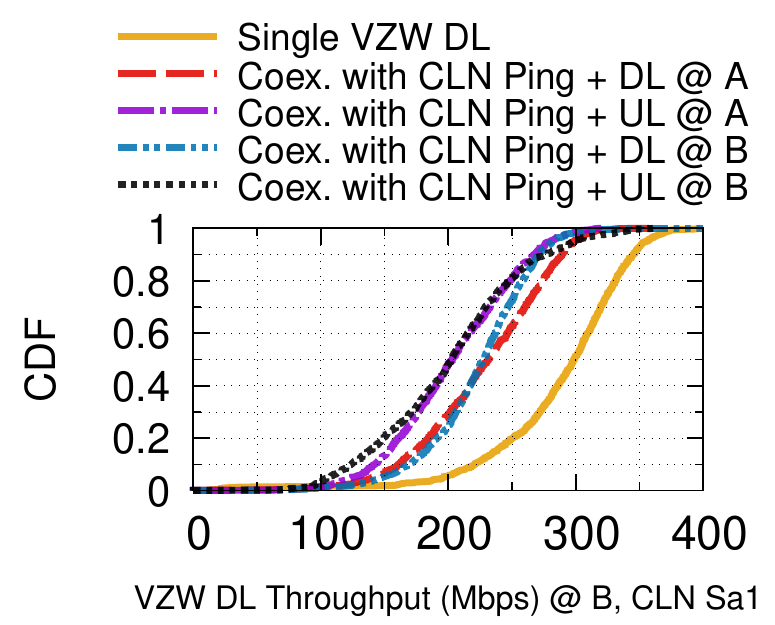}
    \caption{\vzw DL tput. @ B, \cln Sa1}
    \label{fig:tputVzwDlBSa1Gap}
    \end{subfigure}
    \hfill
    \begin{subfigure}{.245\textwidth}
    \includegraphics[width=\linewidth]{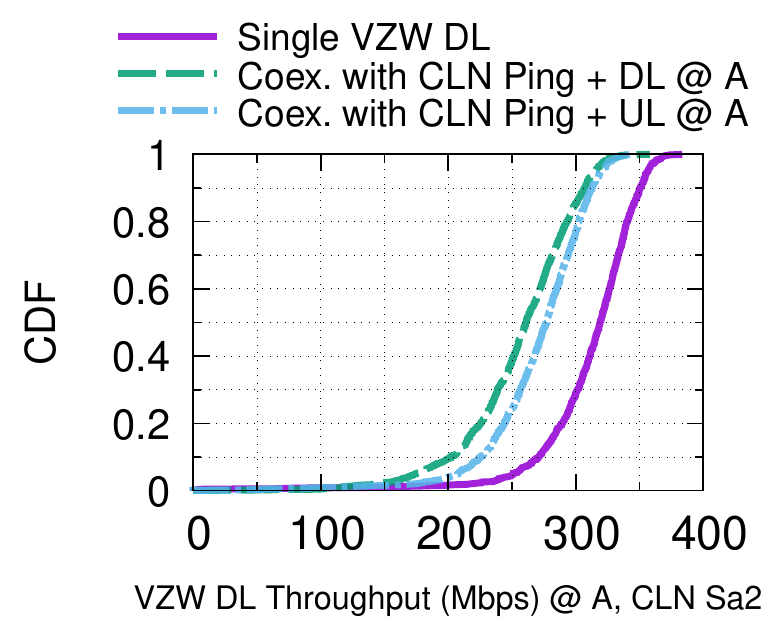}
    \caption{\vzw DL tput. @ A, \cln Sa2}
    \label{fig:tputVzwDlASa2Gap}
    \end{subfigure}
    \hfill
    \begin{subfigure}{.245\textwidth}
    \includegraphics[width=\linewidth]{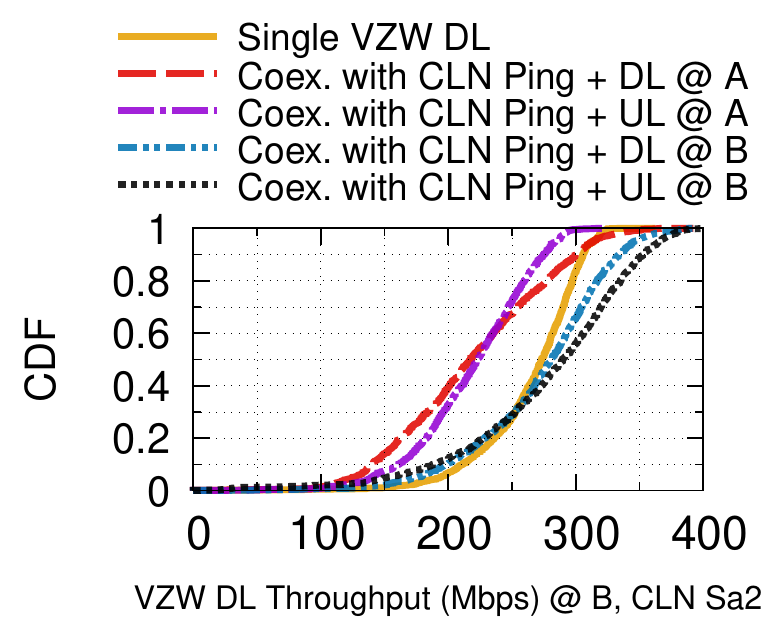}
    \caption{\vzw DL tput. @ B, \cln Sa2}
    \label{fig:tputVzwDlBSa2Gap}
    \end{subfigure}
    \caption{Coexistence performance in terms of \vzw DL throughput with 20 MHz guard band.}
    \label{fig:tputVzwDlTddGap}
    \vspace{-1em}
\end{figure*}

\begin{figure*}[htb!]
    \centering
    \begin{subfigure}{.325\textwidth}
    \includegraphics[width=\linewidth]{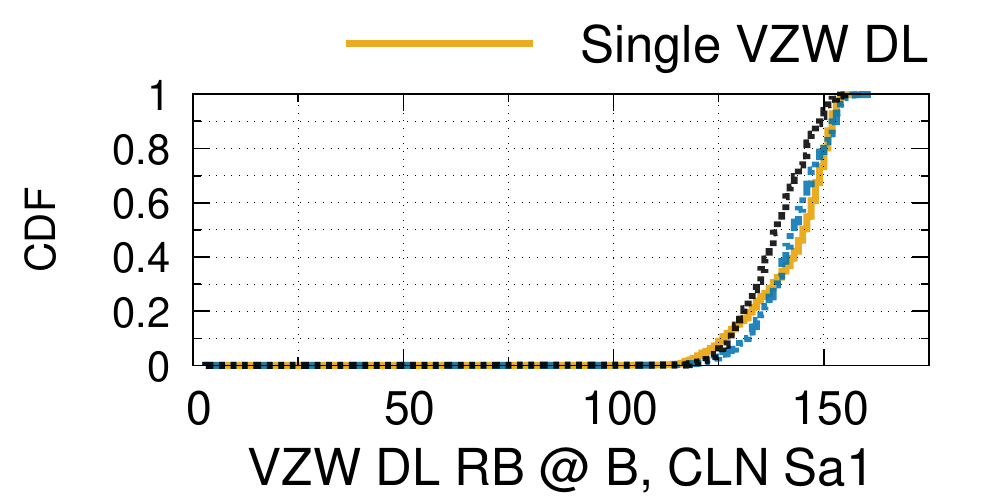}
    \caption{\vzw DL RB}
    \label{fig:tputVzwDlARbSa1Gap}
    \end{subfigure}
    \begin{subfigure}{.325\textwidth}
    \includegraphics[width=\linewidth]{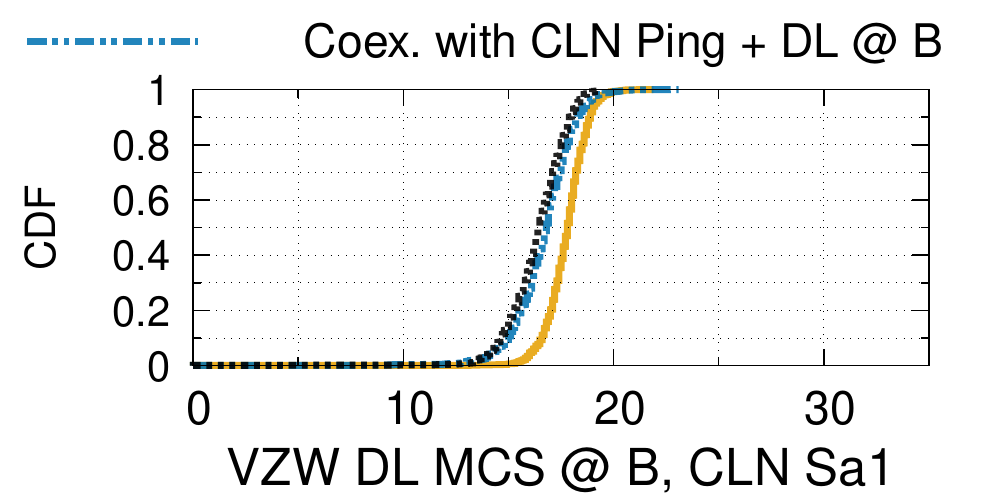}
    \caption{\vzw DL MCS}
    \label{fig:tputVzwDlAMcsSa1Gap}
    \end{subfigure}
    \begin{subfigure}{.325\textwidth}
    \includegraphics[width=\linewidth]{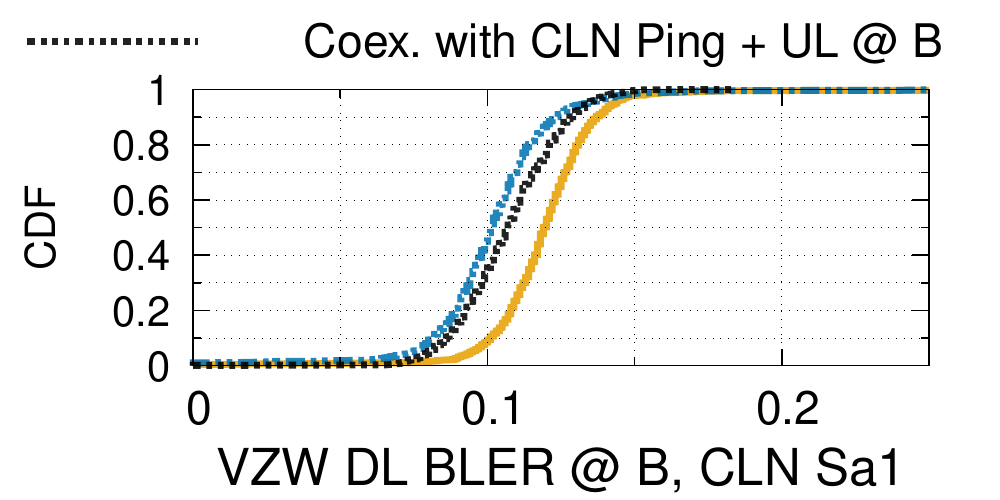}
    \caption{\vzw DL BLER}
    \label{fig:tputVzwDlABlerSa1Gap}
    \end{subfigure}
    \caption{Representative comparison of DL RB, MCS, and BLER for \vzw @ A, on \cln Sa1, with 20 MHz guard band.}
    \label{fig:tputVzwTdd2Gap}
    \vspace{-1.5em}
\end{figure*}

Fig.~\ref{fig:tputClnDlTddGap} shows the DL throughput performance of \cln under \textbf{GAP} scenarios. When \cln uses Sa1 configuration (Fig.~\ref{fig:tputClnDlASa1Gap}, \ref{fig:tputClnDlBSa1Gap}), we observe no throughput degradation due to the low number of DL subframes utilized. When \cln is using Sa2, we observe the highest degradation of 21\% when coexisting with \vzw \ul @ A. This is an improvement from the highest degradation 60\% in the same scenario without the guard band. Further, Fig.~\ref{fig:tputClnTdd2Gap} shows the DL RB, MCS, and BLER of the representative \textbf{GAP} results of \cln and \vzw @ A, when \cln is using Sa2. While the RB allocations stays at the maximum, we observe higher MCS allocation and lower BLER compared to representative \textbf{non-GAP} results on Fig.~\ref{fig:tputClnTdd2}.

Next, Fig.~\ref{fig:tputVzwDlTddGap} shows the DL throughput performance of \vzw under \textbf{GAP} scenarios. We observe throughput degradation on various parameters, with the highest reduction of 30\% when \vzw is coexisting with \cln \ul @ B using Sa1. However, these reductions can be explained by the higher MCS used on the single cases. As a representative result, Fig.~\ref{fig:tputVzwTdd2Gap} shows the DL RB, MCS, and BLER of \textbf{GAP} results of \cln and \vzw @ B, when \cln is using Sa1. We observe a higher MCS and correspondingly, a slightly higher BLER on the single case. Additionally, we observed a median DL BLER of 0.1-0.12 on all cases. Therefore, these throughput degradations are not caused by interference, but network variations.

\section{Conclusions and Future Work}
\label{sec:conclusions}
\vspace{-0.1cm}
This paper presents the first comprehensive measurement-based analysis of the effect of mutual OOB interference on throughput and  latency between CBRS and C-band when the two systems are deployed in adjacent channels with and without a guard band, using both spectrum analyzer based power measurements and detailed throughput and error analyses. It is clear that the combination of no guard bands, power difference and lack of TDD synchronization pose obstacles in attaining the high throughputs expected of both CBRS and C-band deployments. When a 20 MHz gap is added between the CBRS and C-band channels, we observe a reduction in throughput degradation: from 60\% degradation to 21\% on CBRS and from 43\% to 30\% on C-band: this clearly indicates the impact of OOB interference.

There are many potential solutions that can minimize the effect of coexistence between CBRS and C-band: (i) having a static or dynamic guard band based on interfered resource block allocation, (ii) reducing the C-band transmission power on resource blocks that are adjacent to the CBRS frequency, and (iii) common TDD configuration between C-band and CBRS. Some of these solutions will require changes to the standard. We plan to study the above solutions with analysis and simulations, and further experiments using software defined radios to capture raw I/Q data on both systems for further detailed analysis on interference profiles in the frequency domain. We also plan to study coexistence scenarios that we have not yet explored, \eg coexistence between outdoor CBRS and outdoor C-band, and between 5G CBRS and 5G C-band. 

\vspace{-0.5em}
\section*{Acknowledgment}
\vspace{-0.1cm}
We gratefully acknowledge the Facilities office at the University of Chicago for providing access to the Harper Court building.
\vspace{-0.5em}

\bibliographystyle{IEEEtran}
\bibliography{IEEEabrv,references}

\end{document}